\documentclass[preprint]{elsarticle}
\usepackage{ifpdf}
\usepackage{amssymb}
\usepackage{epsfig,graphicx}
\usepackage{graphicx}
\usepackage{amsmath}
\usepackage{color}

\newcommand{\be}{\begin{equation}}
\newcommand{\ee}{\end{equation}}

\newcommand{\bee}{\begin{eqnarray}}
\newcommand{\eee}{\end{eqnarray}}

\def\be{\begin{eqnarray} &&}

\def\ee{\end{eqnarray}}
\def\bew{\begin{widetext}}
\def\ew{\end{widetext}}

\usepackage{epsfig}
\def\lp {\left( }
\def\rp {\right) }
\def\lb {\left[ }
\def\rb {\right] }
\def\lc {\left\{ }
\def\rc {\right\} }

\def\bea{\begin{eqnarray}}
\def\eea{\end{eqnarray}}
\def\nn {\nonumber}

\def\D {\Delta}

\def\p {\pi}

\newcommand{\bppp}{B^+ \to  \pi^-\pi^+\pi^+}
\newcommand{\bkkk}{B^+ \to  K^-K^+K^+}

\begin{document} 
\makeatletter
\begin{frontmatter}
\title{Charm Penguin in $B^\pm \to K^\pm K^+ K^-$: partonic and hadronic loops}
\author[CBPF]{I. Bediaga} 
\author[ITA]{T. Frederico}
\author[CBPF]{P. C. Magalh\~aes}
\ead{pmagalhaes@cbpf.br}

\address[CBPF]{Centro Brasileiro de Pesquisas F\'isicas, \\
22.290-180, Rio de Janeiro, RJ, Brazil}
\address[ITA]{Instituto Tecnol\'ogico de
Aeron\'autica, DCTA \\ 12.228-900 S\~ao Jos\'e dos Campos, SP,
Brazil.}
\date{\today}
\begin{abstract}
Charm penguin diagrams are known to be the main contribution to charmless B decay process with strangeness variation equal to minus one, which is the case of $B^\pm \to K^\pm K^+ K^-$ decay. The large phase space available in this and other B three-body decays allows non trivial final state interactions with all sort of rescattering processes and also access high momentum transfers in the central region of the Dalitz plane. In this  work we investigate the charm Penguin contribution to $B^\pm \to K^\pm K^+ K^-$, described by a hadronic triangle loop  in nonperturbative regions of the phase space, and by a partonic loop at the quasi perturbative region. 
These nonresonant amplitudes  should have  a particular structure in the Dalitz plane and their contributions to the final decay amplitude can  be confirmed by a data amplitude analysis in this channel. 
In particular, the hadronic amplitude has a changing sign in the phase at $D\bar{D}$ threshold which can result in a change of sign for the CP asymmetry. 
\end{abstract}
\end{frontmatter}

{\it Introduction.} The general method  to access directly CP asymmetries and partial branching fraction in charmless B decays uses mainly  the relative contributions of Penguins and Trees quark diagrams. In the BSS~\cite{BSS} approach  the weak phase comes from the Tree amplitude, which interferes with the strong phase coming from the Penguin amplitude producing CP violation. 
The factorization approach within this method describes  well the two-body charmless  B decay branching fraction~\cite{QCDF}. However, the same is not true for the predicted CP asymmetries, where there are several deviations from the experimental data~\cite{pdg}. 

 The factorization approach has been also used  for charmless three-body B decays, although, in this case, it  is a more delicate approximation. The form factors present in these three-body decays are much more complex,  {\it depending on two Dalitz variables} and spread through the large energy range available in these decays. In general they are parametrized by resonances, based in the quasi two-body approximation for the decay process.
  The nonresonant contribution is a complicated issue: the full treatment should include proper three-body  rescattering 
  effects which are  not well understood. From the experimental analysis side, they usually fit data with ad hoc functions that are not based in any fundamental or phenomenological theory. On 
  the other hand, the authors in Refs.~\cite{Fajfer,chinos} used   Heavy Meson Chiral Perturbation Theory (HMChPT) to  estimate
  nonresonant 
form factors in $B\to hhh$ ($h\equiv$ light mesons) and argued that they  are dominated by tree quark topologies.
However, these amplitudes are limited 
to  kinematic regions where the two-body invariant mass of the pair in the final state is small enough to validate ChPT.

 When moving to  hadronic (long distance) interaction contributions in  charmless  three-body B decays, two  out of the three 
 light pseudo-scalars  in the final state have access  
to a large range of energy  in the available phase space, which allow them to rescatter into other mesons. Although    
absent in factorization approach,  many authors \cite{wolfenstein,Gerard,Soni2005,BOT,ABAOT,GLR}  
 have shown that rescattering plays an important role in B decays. 
  In particular, they proved the relevance for B two-body charmless decays
 of charm mesons rescattering into light ones, namely, in the understanding of the observed Branching 
 fractions~\cite{Soni2005,GLR} and CP violation~\cite{wolfenstein,Gerard,Soni2005}. It is remarkable  that this
  rescattering  contribution was never  studied before within a three-body formulation.

\begin{figure}[ht]
\begin{center}
\hspace*{-4mm}\includegraphics[width=.35\columnwidth,angle=0]{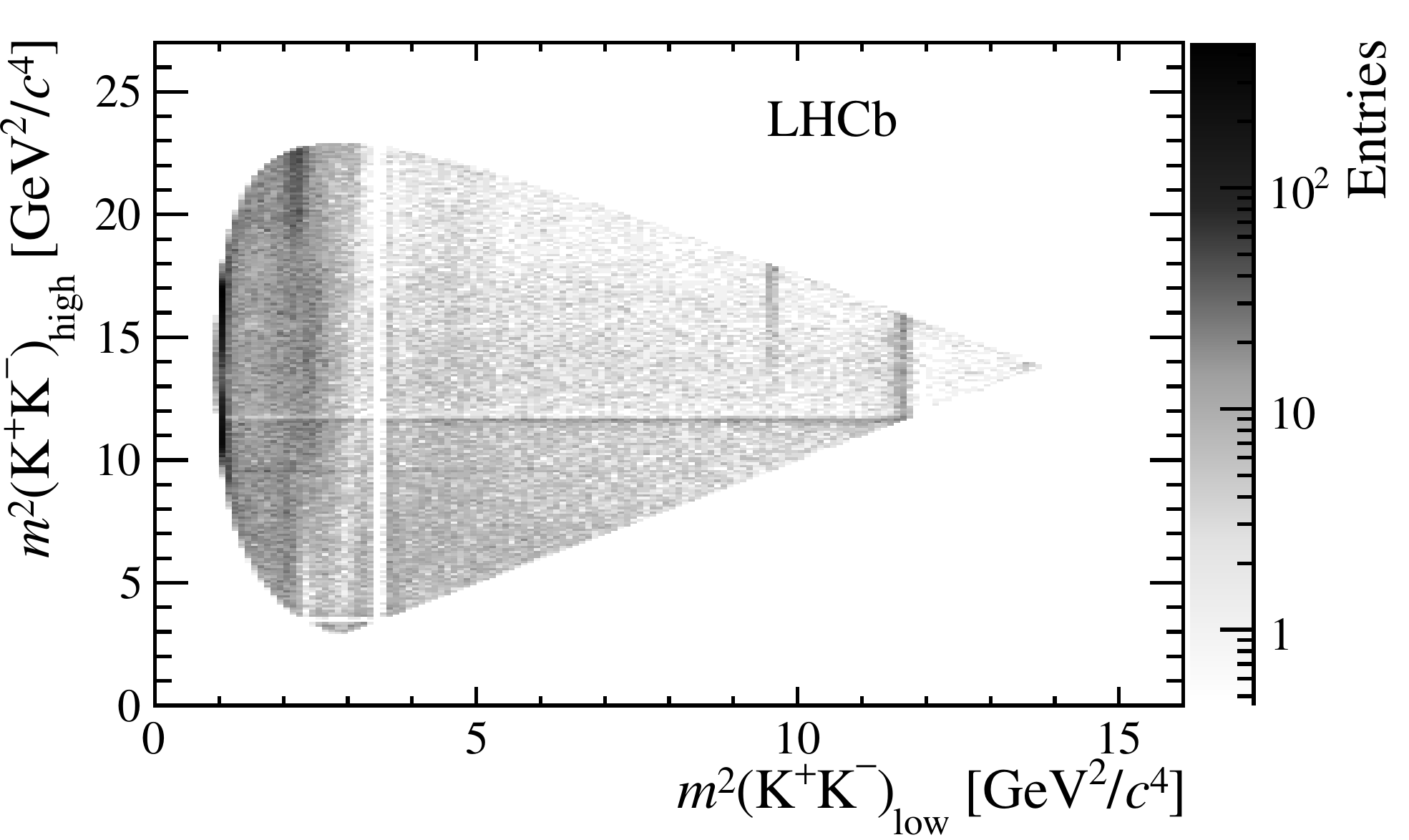}
\hspace*{-2mm}\includegraphics[width=.34\columnwidth,angle=0]{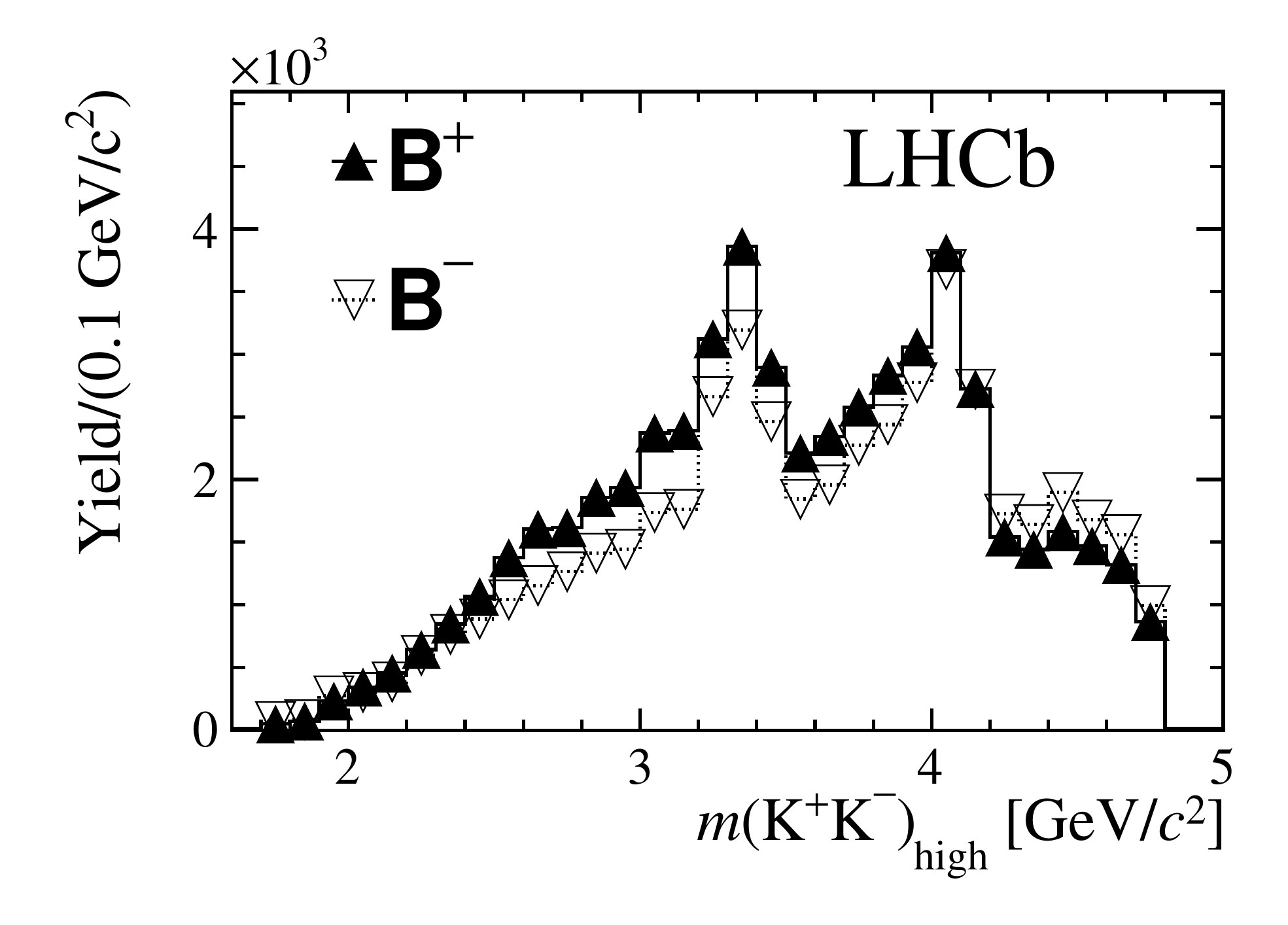}
\hspace*{-2mm}\includegraphics[width=.34\columnwidth,angle=0]{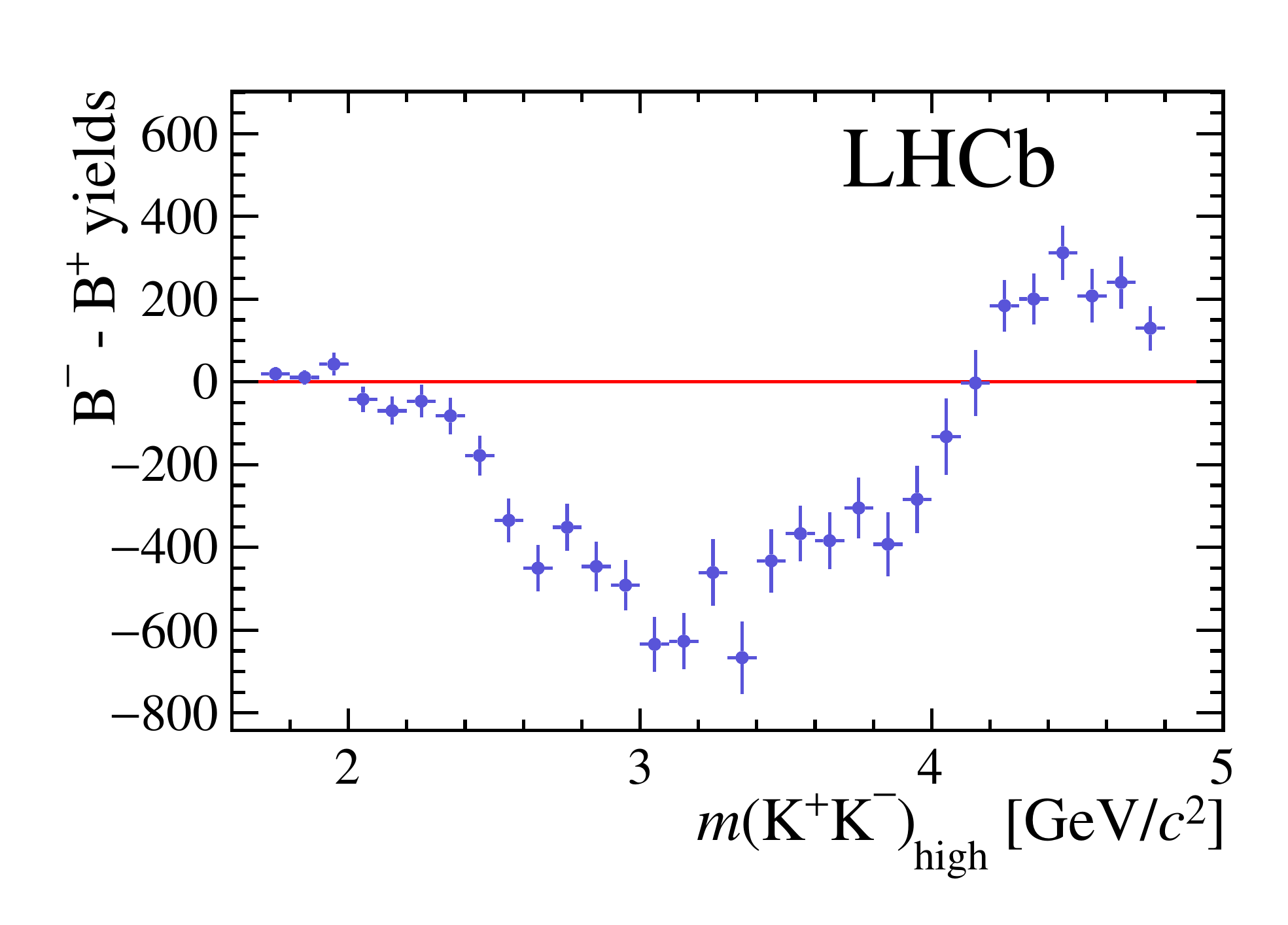}
\caption{  $B^\pm \to K^\pm K^+ K^-$ decay from LHCb experiment~\cite{LHCbPRD2014}: (left) full data Dalitz plot ($B^+ + B^-$); (center) events for $B^+$ and $B^-$ projected on $m(KK)_{high}$ \cite{LHCb_addInfos}  ; and (right) the CP-Asymmetry ($B^+ - B^-$) of the events projected on $m(KK)_{high}$\cite{LHCb_addInfos}.}
\label{lhcb}
\end{center}\end{figure}

In this paper we study the contribution of a double charm intermediate interaction to the  $B^\pm \to K^\pm K^+ K^-$ decay.  
  Although this process has some suppression, the weak decay involving two charm quarks is more favourable than the one with
two light quarks, which can compensate this suppression and give a significant  contribution to the total decay amplitude. 
The $B^\pm \to K^\pm K^+ K^-$ process is a particular interesting place to study this contribution because: 
(i) it has a large BR compared to  other charmless three-body B decays;
(ii) it is dominated by the penguin weak topology; and
(iii) the experimental data from LHCb \cite{LHCbPRD2014}, Fig. ~\ref{lhcb}(left), show a significant
 population of events spread up to high values of invariant masses, confirming previous data distribution from BaBar~\cite{BabarCP} and Belle\cite{Belle}  on this channel.

  The same LHCb paper~\cite{LHCbPRD2014} study the CP asymmetry distribution in the Dalitz plot for the four channels:  $B^\pm \to K^\pm \pi^+\pi^-$, $B^\pm \to \pi^\pm \pi^+\pi^-$, $B^\pm \to \pi^\pm K^+ K^-$, $B^\pm \to K^\pm K^+K^-$. In particular, they showed  a clear correlation  
  between the channels $B^\pm \to K^\pm \pi^+\pi^-$ and  $B^\pm \to K^\pm K^+ K^-$ decays,  observed in the region where 
   $\pi^+\pi^-\to K^+K^-$ has an important contribution in the hadron-hadron scattering amplitude~\cite{Cohen} 
   - i.e. between 1 and 1.6 GeV. The $B^\pm \to K^\pm \pi^+\pi^-$ has a positive CP asymmetry in this region whereas the $B^\pm \to K^\pm K^+ K^-$ has a negative one. 
 A similar correlation in the CP asymmetry, i.e. in the same mass region, was observed between the two channels 
 $B^\pm \to\pi^\pm K^+K^-$ and $B^\pm \to \pi^\pm \pi^+\pi^-$. These results indicate that the rescattering process  
 $\pi^+\pi^-\to K^+K^-$ is present in these decays~\cite{BOT,ABAOT}, carrying the strong phase necessary for CP violation 
 and conserving CPT global symmetry as discussed in Ref.~\cite{BOT,ABAOT}. 

The Fig.~\ref{lhcb}(center)  shows  the events for $B^+$ and $B^-$ integrated in $m(KK)_{low}$ presented by LHCb ~\cite{LHCb_addInfos} 
for the $B^\pm \to K^\pm K^+ K^-$ decay, where the two peaks corresponds to the vector resonance $\phi(1020)$ in this particular projection. By subtracting both curves in  Fig.~\ref{lhcb}(center) we access the amount of events related to CP violation on that  projection, Fig.~\ref{lhcb}(right).  Inspecting  Fig.~\ref{lhcb}(right) it is possible to identify that the negative CP asymmetry is placed in the region where the rescattering $\pi\pi \to KK$ we mention above is important in the $m(KK)_{low}$ variable. After that, the CP asymmetry changes sign crossing zero  at $~ 4$ GeV, near the $D\bar{D}$ open channel. Moreover, LHCb~\cite{LHCb_addInfos} data distribution observes the same change in CP asymmetry sign at 4 GeV in  $B^\pm \to K^\pm \p^+ \p^-$ but with an opposite direction.  The same correlation was also observed between the channels $B^\pm \to \p^\pm \p^+ \p^-$ and $B^\pm \to \p^\pm K^+ K^-$ at the same 4 GeV invariant mass.   
Analogously of what was seen for the $\pi^+\pi^-\to K^+K^-$ rescattering contribution to  three-body charmless B decays,
 we  investigate the hypotheses that the 
rescattering  process $D\bar{D}\to P\bar{P}$ could provide also the strong phase  needed to observe CP asymmetry 
in the high mass region.

{\it Charm Penguin Dynamics.}  In a recent paper \cite{ManeG}, the authors   discussed the characteristics of the three-body momentum distribution along the phase space,  for the particular process $\bppp$. They showed that the peripheral regions of the Dalitz plot, where the light resonance is placed, are essentially nonperturbative.
On the other hand, the central region of the Dalitz is dominated by large transfer momentum requiring a quasi perturbative treatment of QCD.

\begin{figure}[ht]
\begin{center}
\includegraphics[width=.45\columnwidth,angle=0]{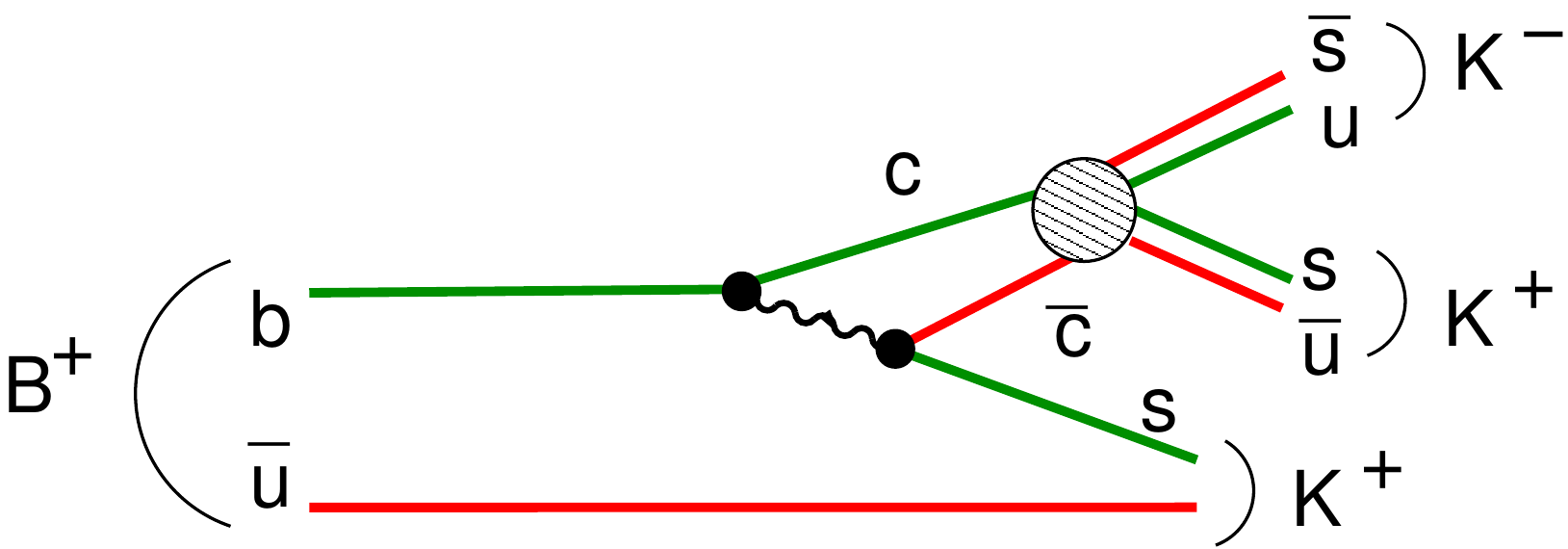}
\caption{ Penguin weak topology diagram  for $B^\pm \to K^\pm K^+ K^-$. }
\label{fig:penguin}
\end{center}
\end{figure}

Within this scenario the charm Penguin (CharmP) diagram, in Fig.~\ref{fig:penguin}, contributes in distinct Dalitz 
regions with a different behaviour: one involving short distance  physics expressed by partons loop and placed in central region; 
and the  other one involving  the long distance dynamics, which can be described by hadron loops, and are expected to 
be relevant in the peripheral Dalitz region. 
 Other than give a significant contribution for the total decay rates,  the  CharmP can be the mechanism to explain  experimental observations in charmless three-body  B  decays: the abundant phenomena of CP violation at high masses,  providing  the strong phase one needs; and 
 the significant population of the high mass phase space by a nonresonant amplitude.

\begin{figure}[ht]
\begin{center}
\includegraphics[width=.3\columnwidth,angle=0]{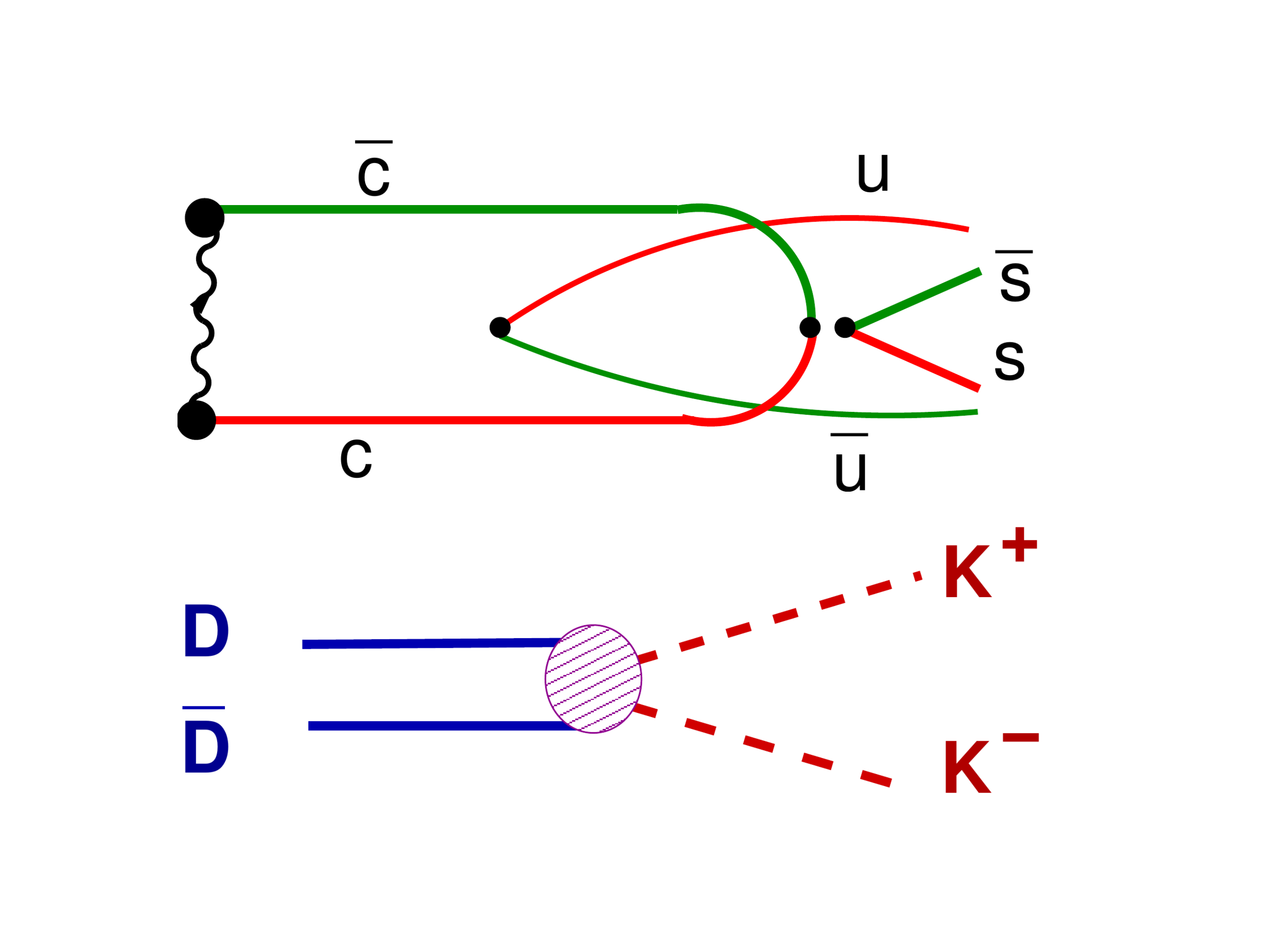}
\includegraphics[width=.4\columnwidth,angle=0]{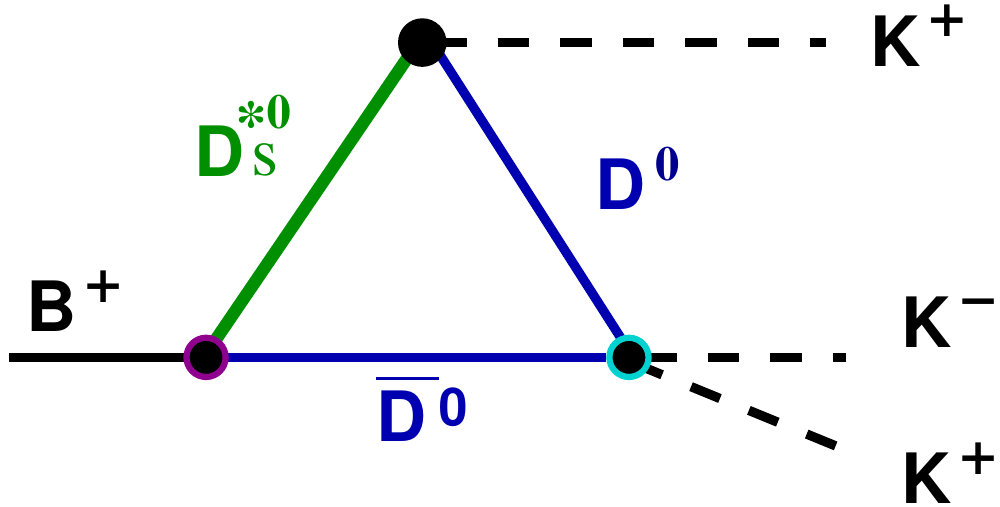}
\caption{ Left diagrams: double charm partonic loop producing $K^+K^-$ (upper panel) and
double charm hadronic loop producing $K^+K^-$ (lower panel). Right: triangle diagram for hadronic loop for $\bkkk$ with vector form factor. }
\label{fig:weakH}
\end{center}\end{figure}

In order to check to which extend the separation between short and long distance can be used to represent the $B^\pm \to K^\pm K^+ K^-$ decay amplitude, 
 we investigate the two Charm Penguin contributions at the partonic and  mesonic levels represented, respectively, 
 in figures \ref{fig:penguin} and  \ref{fig:weakH}.  
 The kinematical range where these contributions may be dominant are studied  and we found 
 quite different patterns for the  two Charm Penguin contributions at the partonic and at the meson levels. We study
 their signatures and contributions to the final decay amplitude that should be identified in a future amplitude data analyses.

{\it Partonic Charm Penguin.} B decays involving strangeness variation equal to minus one are dominated by the Penguin contribution, 
which is the case of the $B^\pm \to K^\pm K^+ K^-$ decay. Inspecting the LHCb data\cite{LHCbPRD2014} in Fig.~\ref{lhcb}(left) one can note that in the middle
 of the Dalitz plot, 
i.e. the  region where we could expect  partonic physics to play an important role, is populated
 with a considerable number of events.  
Moreover, in the same region, the data shows the  undoubted presence of the scalar $\chi_{c0}(3415)$, which is also a hint that this 
is a rich $c\bar{c}$ environment for the nonresonant scalar amplitude from the charm penguin to take place.

\begin{figure}[ht!!!!]
\begin{center}
\includegraphics[width=.45\columnwidth,angle=0]{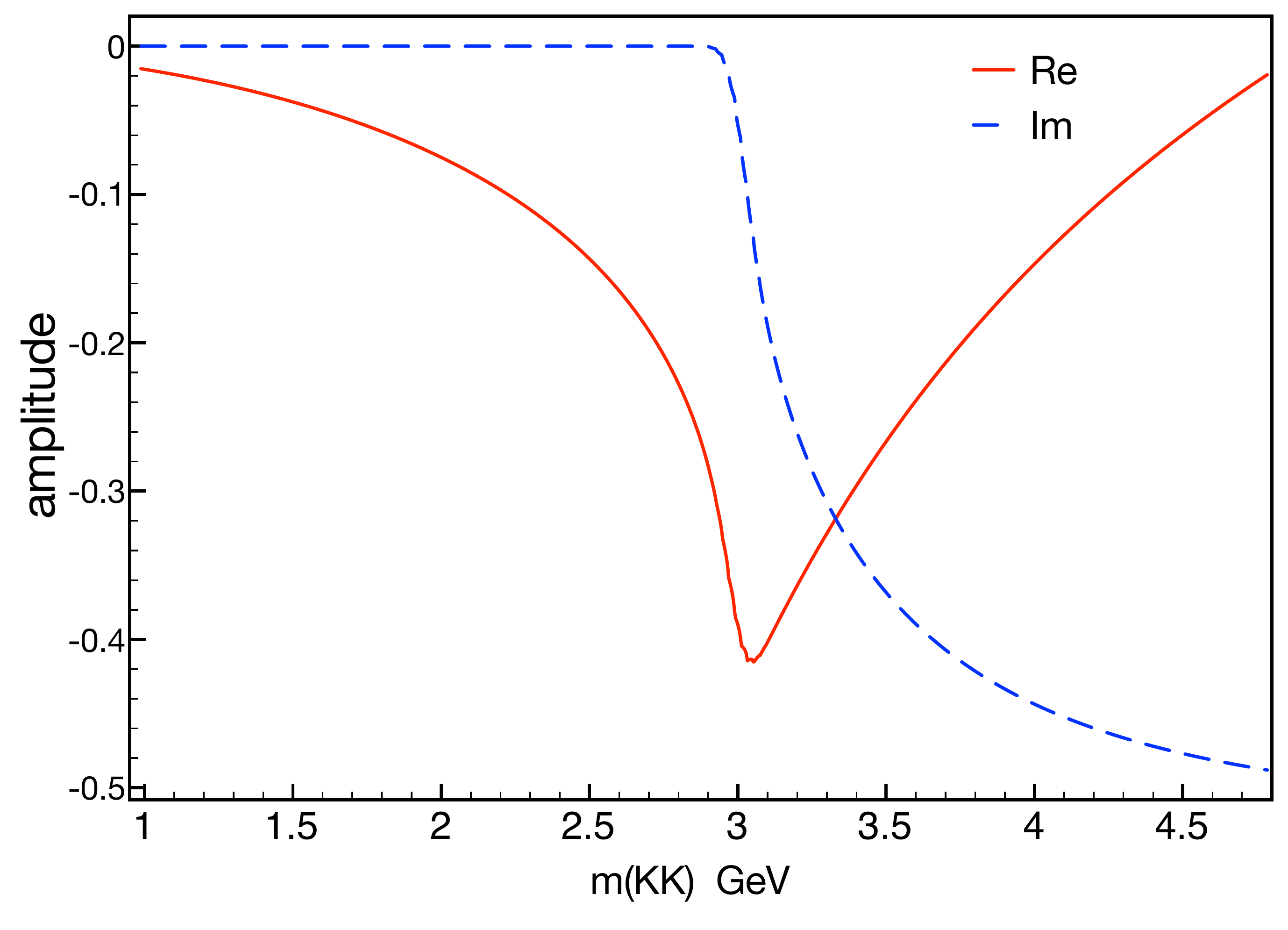}
\includegraphics[width=.45\columnwidth,angle=0]{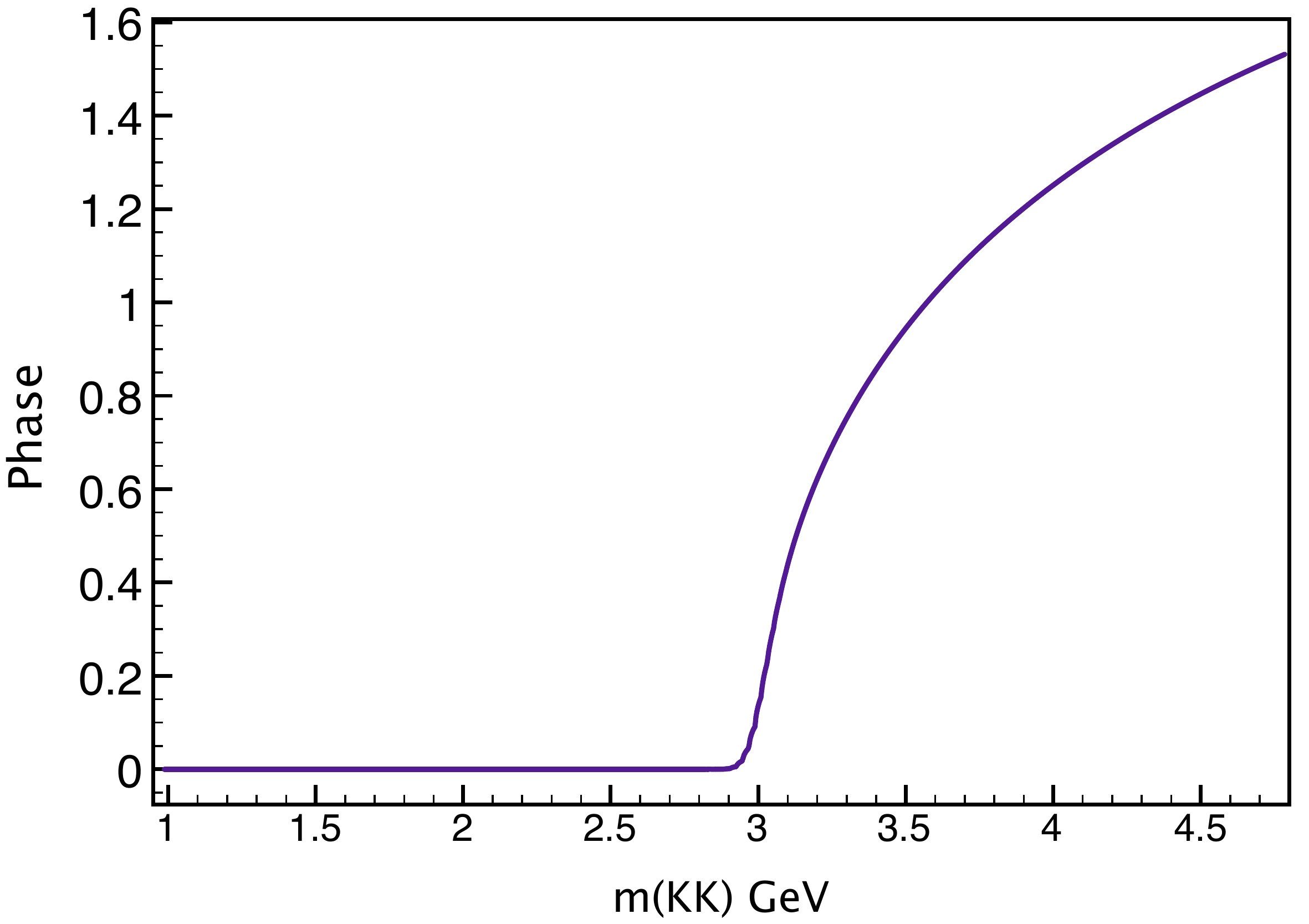}
\caption{Partonic charm Penguin proposed by Ref.\cite{Gerard}, eq.(\ref{gerard}): (left) real (dispersive), imaginary(absorptive) amplitudes; (right) phase in radians. }
\label{fig:MagPhaseGerard}
\end{center}
\end{figure}

 We considered the charm penguin contributions as represented by the diagram of Fig. \ref{fig:penguin}.
 However, is very hard to precise  the effective charm mass propagating inside the loop due to the exchange of gluons 
 and how the hadronization affects this picture. 
To guide our calculation one follows the structure proposed by Mannel et al. \cite{ManeG} to describe the center region of the 
Dalitz plot for $\bppp$.
   The authors propose a functional form of this amplitude to be
  $A_p (s) = T(s)( M_B^2 - s ) f_+(s )$.
 Translating to  $\bkkk$ process, $f_+(q^2)$ is the $B\to K$ vector form factor, which can assume the single pole parametrization:
$f_+(s) = \frac{1}{1 - s/ M_{Bs}^{*2} }$,
 with $M_{Bs}^*$ being the mass of a vector 
meson $B_s^*$. The function $T(s)$ is the kernel, which we identify as the charm parton loop. 
The $c\bar{c}$ bubble loop contribution is very well known and was calculated also by  Gerard and Hu (1991)\cite{Gerard}, with a real 
and imaginary part given by: 
 \begin{eqnarray}
 \Re\Pi(x) &=& -\frac{1}{6} \lc \frac{5}{3} + \frac{4}{x} - \lp 1+\frac{2}{x}\rp \lb \sqrt{1-\frac{4}{x}}\,\, \ln\lp\frac{1+\sqrt{1-4/x}}{1- \sqrt{1-4/x}}\rp \Theta\lb 1- \frac{4}{x}\rb  \right. \right.\nn\\+&& \left.\left. 2\, \sqrt{\frac{4}{x} - 1}\,\, \cot^{-1}\lb \sqrt{\frac{4}{x} - 1}\rb \Theta\lb  \frac{4}{x}-1\rb 
  \rb \rc\,, \nn \\[2mm]
  \Im\Pi(x) &=&-\frac{\pi}{6}\,\lp 1+\frac{2}{x}\rp\, \sqrt{1-\frac{4}{x}}\, \Theta\lb 1- \frac{4}{x}\rb \,,
  \label{gerard}
 \end{eqnarray}   
 where $x= s/m_c^2$. In Fig.~\ref{fig:MagPhaseGerard} one can recognize that the double charm loop behaves exactly as all
  bubble loop function,  which are well known.

The goal here is precise. Once charm mass is about one third of B mass, charm Penguin could give a clear signature in charmless three-body 
B decay. Indeed the effect described by Gerard and Hu in Fig.~\ref{fig:MagPhaseGerard}, i.e. the  maximum of the  real contribution 
 and the beginning of the imaginary contribution,  are  inside the three body  phase space.

\begin{figure}[ht!!!!]
\begin{center}
\includegraphics[width=.45\columnwidth,angle=0]{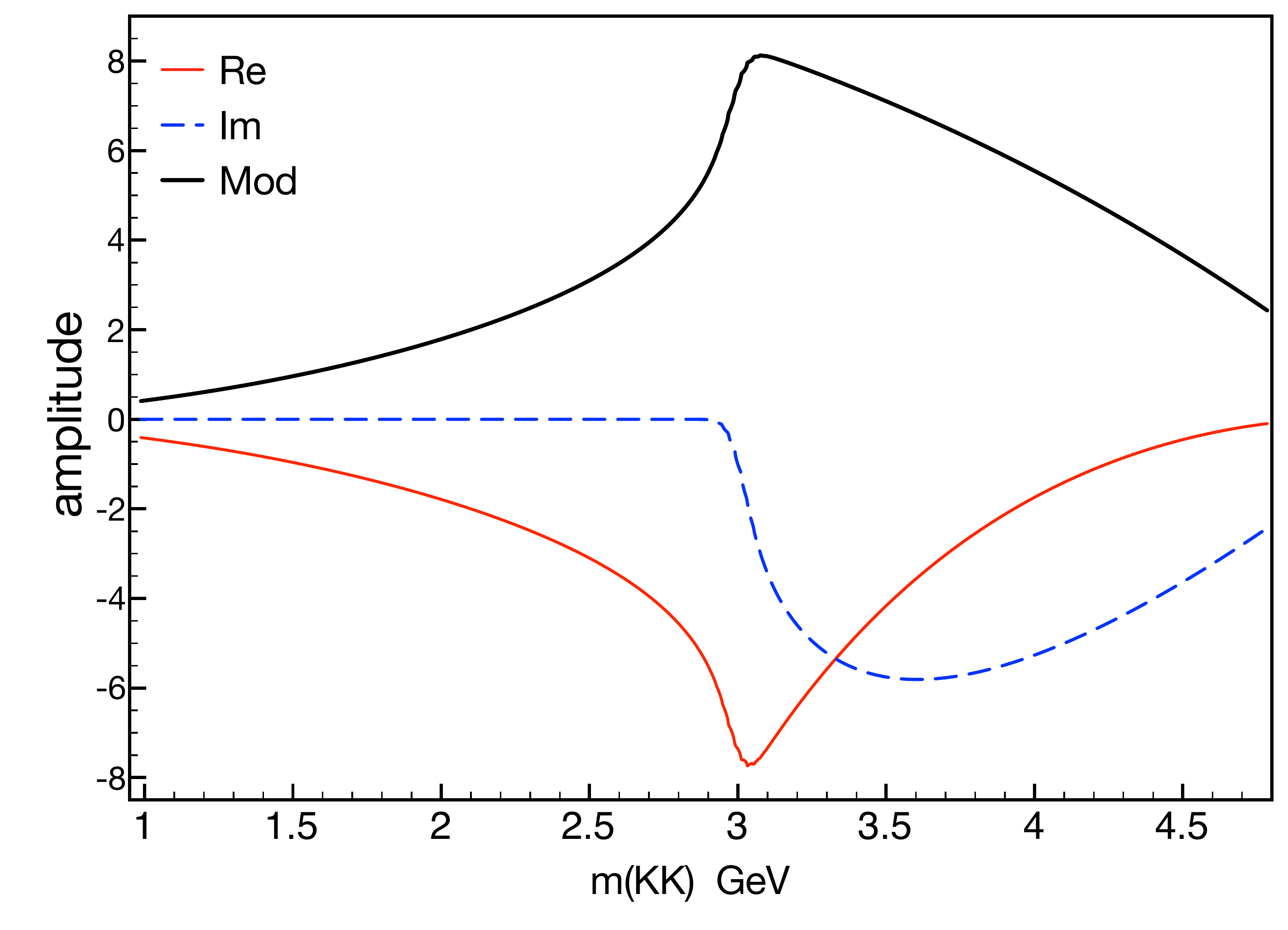}
\includegraphics[width=.45\columnwidth,angle=0]{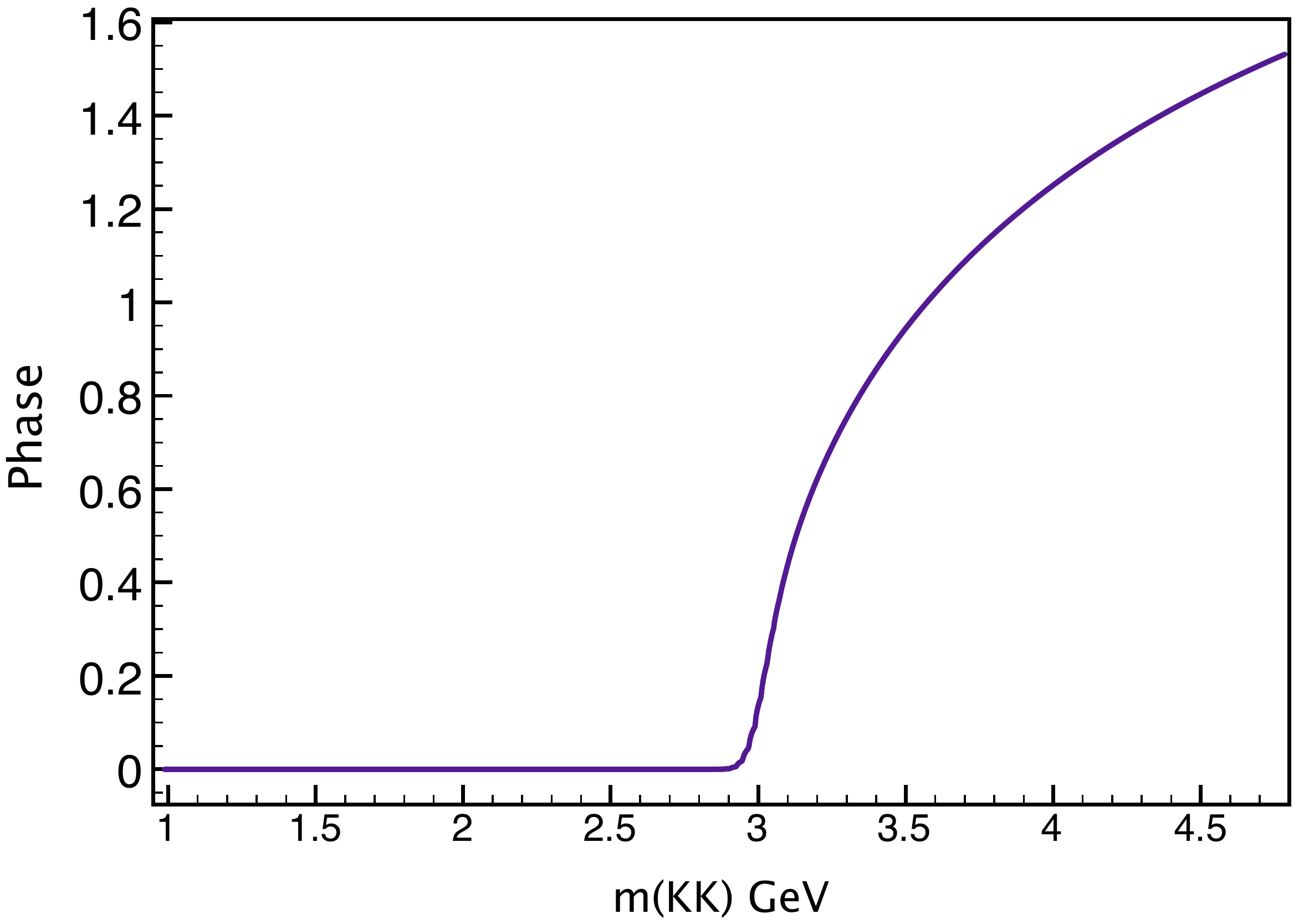}
\caption{ Modulus, real and imaginary parts (left) and phase (right) of the total partonic charm  Penguin amplitude, eq. (\ref{Ap}).}
\label{fig:LoopP}
\end{center}
\end{figure}

As we have discussed previously, the issue on the partonic charm loop is the value of its mass.  
In order to accommodate this uncertainties, we integrate the bubble loop quark function in the charm mass convoluted with a Gaussian 
distribution centred in $m_c =1.5$ GeV and width $\Gamma=20$ MeV. Those values could be taken as a free parameter when fitting real data. 
The final contribution to the partonic amplitude becomes:
\bea
A^P_p = ( M_B^2 - s ) f_+(s ) \int_{m_c^-}^{m_c^+} dm \,\Pi(s) \frac{1}{2\,\pi\,\Gamma^2}\,e^{\frac{ (m -m_{c})^2}{2\,\Gamma^2}}\,,
\label{Ap}
\eea
 where $m_c^\pm=m_{c}\pm 1.0$  GeV .
 The results for the nonresonant partonic penguin amplitude and phase are given in Fig.\ref{fig:LoopP}.
Although the final amplitude has an arbitrary normalization there is  a clear peak around 3 GeV. 
The phase is zero below threshold and rise continues after it. This phase variation will, if present, 
change the  interference pattern with the other amplitudes, which could be noticed in data.

{\it Hadronic Penguin.} The nonresonant  hadronic charm loop is expected to be important for low 
 relative momentum between the mesons in the final state, corresponding to the boundaries of the Dalitz plot. 
Despite of the hadronization effect,  one can expect  the weak transition amplitude to be described by the diagram in the left panel
 of Fig.\ref{fig:weakH}. However, we used an effective description in terms of hadronic degrees of freedom which simplifies these 
 interactions and are summarized by the triangle loop given in the right panel of Fig.\ref{fig:weakH}. It is worth to mention that there could be a superposition of similar processes with excited $D_s^*$ states, but here we are considering only the ground state $D_s^{+*}$ with mass $2.1$ GeV. 

In the triangle loop, one note that besides the weak vertex and the triangle loop itself, we need the scattering amplitude
 $D\bar{D}\to K\bar{K}$, which is not known in literature. Because of the different scales it is difficult to extract this 
 interaction from a fundamental Lagrangian, what would require $SU(4)$~\cite{SU4}. Therefore, we propose a phenomenological 
 amplitude  $T_{D\bar{D}\to K\bar{K}}(s)$ based on S- matrix unitarity and inspired in Regge theory,
  which is developed in details in the appendix \ref{ap:DDKK} (note that this amplitude is concisely denoted by $t_{12}$). 
For the hadronic triangle loop  we use the same technical tools find in Refs. \cite{PRD84,PatWV} developed for the three-body decays  
$D^+\to K^-\p^+\p^+$ and also applied to $B^+ \to \pi^- \pi^+ \pi^+ $ \cite{PatIg}. The weak vertex parameters are inside the constant parameter 
$C_0$ and  the transition matrix $B^+\to D^0 W^+$ is described by a form factor. 

The total amplitude for the
 hadronic loop including the dressing of the ${D\bar{D} \to K\bar{K}}$ vertex by the 
 $T_{D\bar{D} \to K\bar{K}}(s)$ scattering amplitude is given by:
\begin{equation}
A^h_P= i \,C_0 \,T_{D\bar{D} \to K\bar{K}}(s)\,\int \frac{d^4 \ell}{(2\p)^4} \; 
\frac{\,\lp \D_{D^0} +2\,\D_{\bar{D^0}} - \,2\,s + 3\,M^2_\p + M^2_B - l^2 \rp}{\D_{D^0} \,\D_{\bar{D^0}} \,\D_{D*}\;[l^2 - m_{B^*}]} \,,
\label{A1.2}
\end{equation}
where $\D_{D_i} = m^2_{D_i} - s + i \epsilon$ are the meson propagators. 

\begin{figure}[ht]
\begin{center}
\includegraphics[width=.45\columnwidth,angle=0]{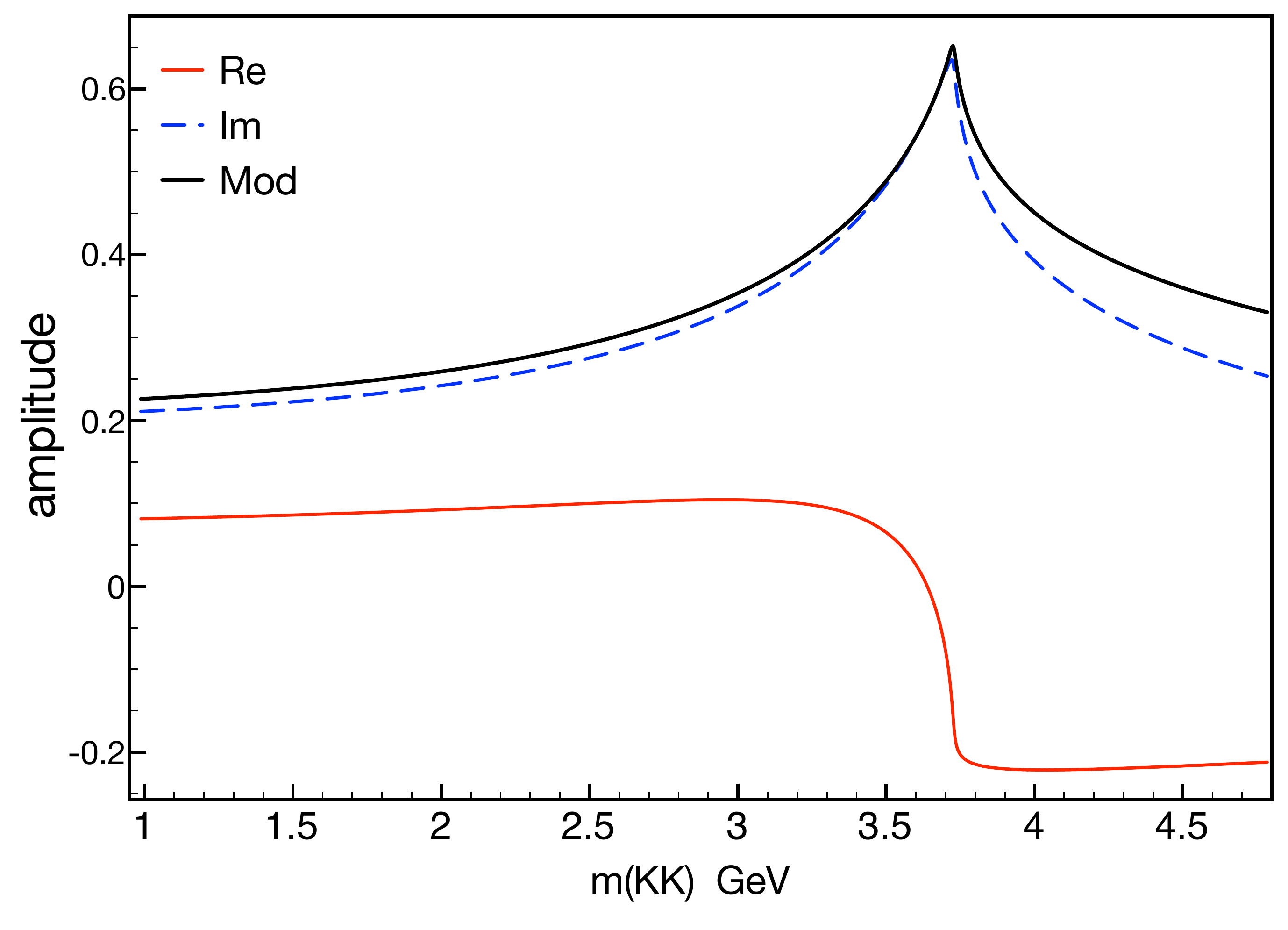}
\includegraphics[width=.45\columnwidth,angle=0]{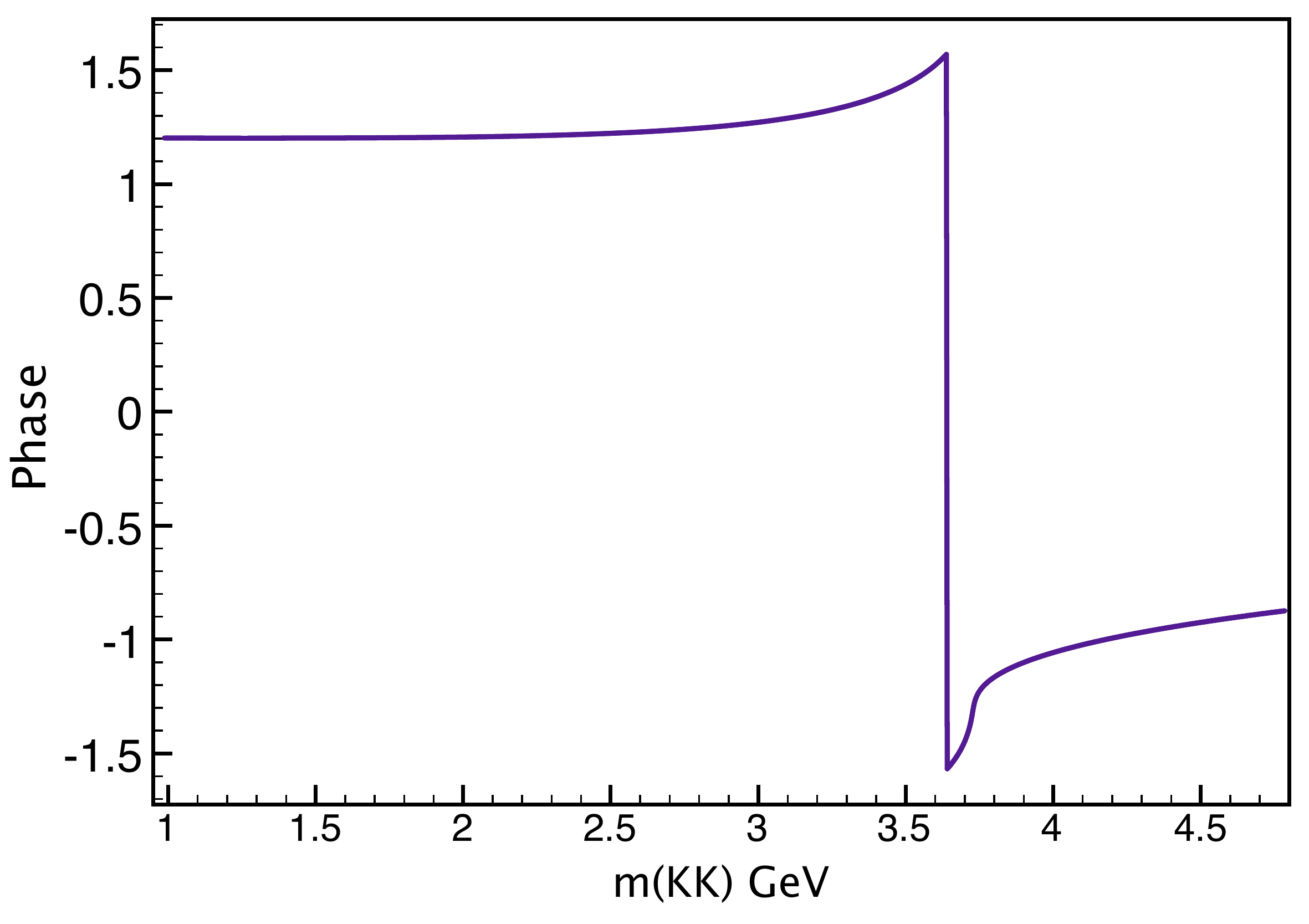}
\caption{ Modulus (left) and phase (right) for the hadronic triangle loop contribution, integral in eq. (\ref{A1.2}). }
\label{fig:Tri}
\end{center}\end{figure}

The exclusive contribution from the hadronic triangle loop, i.e. the integral above,  results in the magnitude and  
phase shown in Fig. \ref{fig:Tri}.
Comparing the results from the hadronic triangle loop, Fig. \ref{fig:Tri}, with  the partonic one, Fig. \ref{fig:MagPhaseGerard},
 one can see that both have a peak at threshold. However, the differences remain on the energy of the open channel and in the 
 absorptive part, which is non zero below the threshold for the hadronic loop.
 
\begin{figure}[ht]
\begin{center}
\includegraphics[width=.45\columnwidth,angle=0]{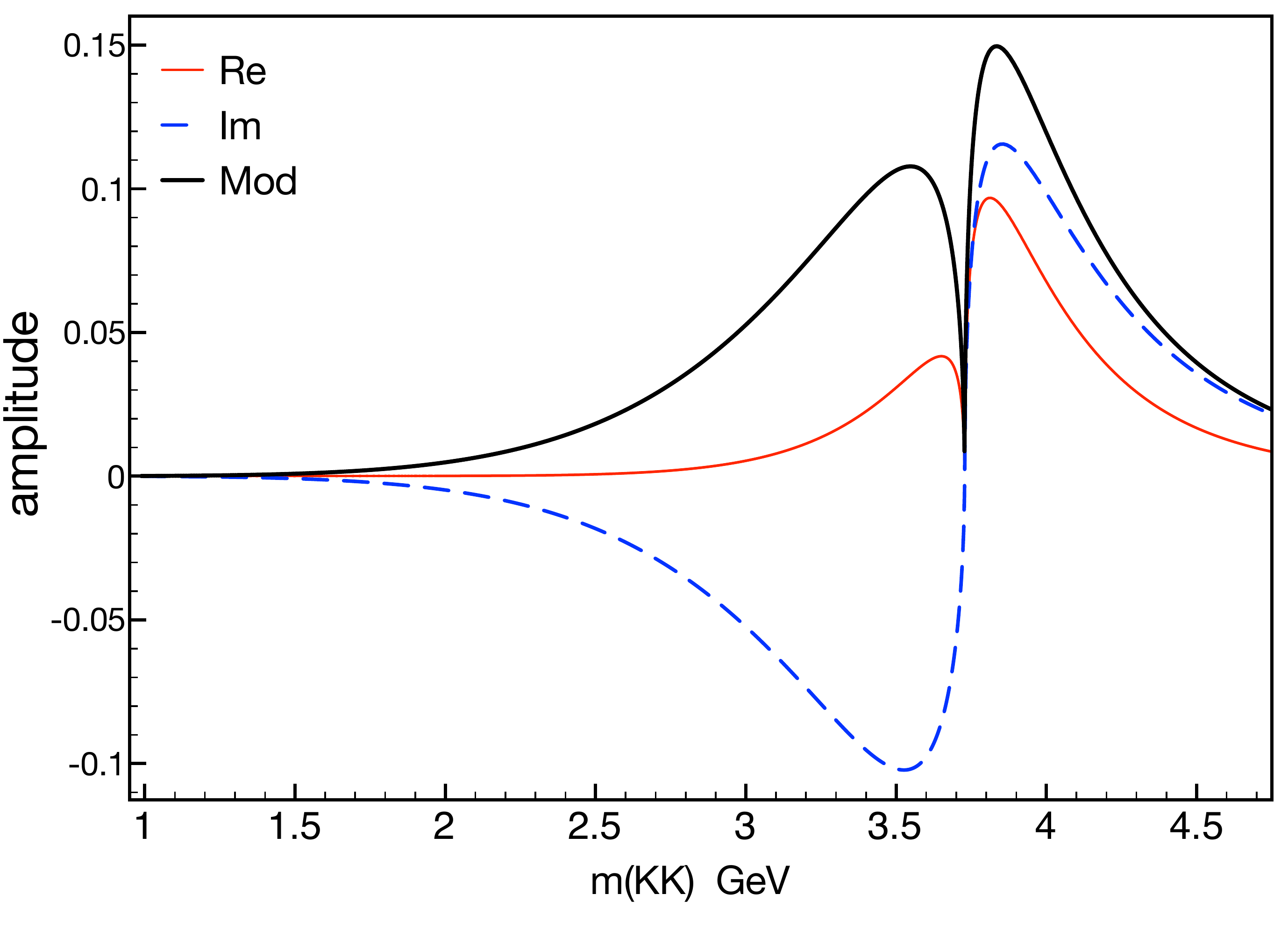}
\includegraphics[width=.45\columnwidth,angle=0]{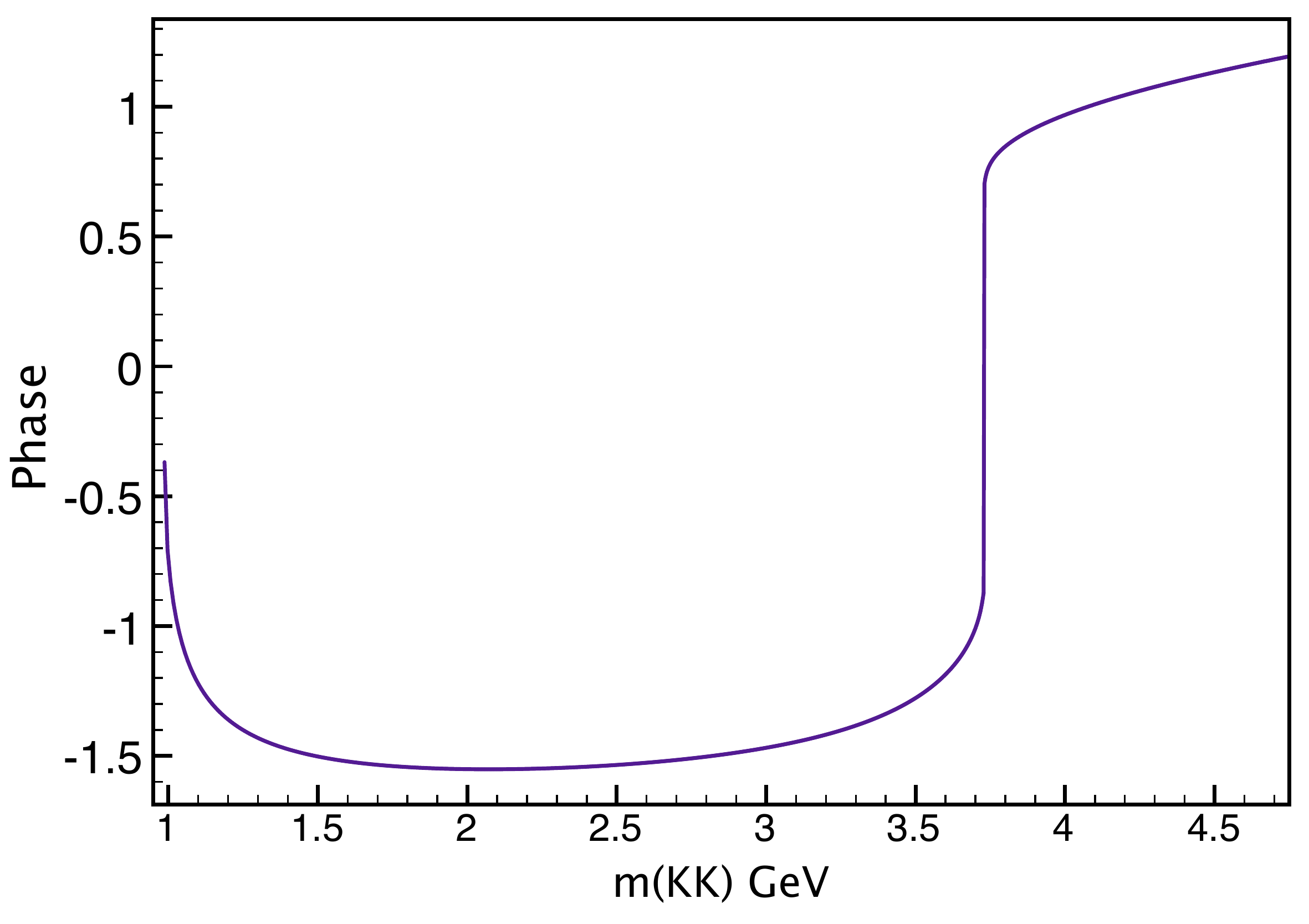}
\caption{Modulus and phase for the total contribution from Hadronic charm penguin, eq. (\ref{A1.2}). }
\label{fig:LoopH}
\end{center}\end{figure}

 The total decay amplitude is obtained after the hadronic loop is multiplied by the $D\bar{D}\to K\bar{K}$ 
scattering amplitude, given by eq. (\ref{t12d}). 
The final results for the magnitude and phase are show in Fig.~\ref{fig:LoopH}. One can note that the rescattering amplitude $D\bar{D}\to K\bar{K}$ plays an important role. It imposes a zero  at the $D\bar{D}$ threshold at the same place the triangle loop has a peak. Although this rescattering amplitude have parameters that needs to be fixed in a fit to data,   the minimum feature  
is that the $D\bar{D}$ threshold  is characterized by a zero between two bumps, with the higher mass one  more pronounced 
and is also where the phase changes it
sign. This changing sign in the phase is a very important characteristic  in order to produce a pattern of interference between amplitudes that leads to changing sign in  CP asymmetry. It is worth remember though that we are considering only one triangle amplitude and the corresponding two-body rescattering into $K\bar{K}$ final state.  

{\it Discussion.}
There are many interesting issues one could explore from our study. 
The structure we follow for the partonic calculation result is wide amplitude which  will be spread in the center of the Dalitz plane. 
 This nonresonant amplitude can explain the significant number 
of events observed in the central region of the Dalitz plot, as show in Fig. \ref{lhcb} (left). 
The hadronic amplitude, on the other side is characterized by two narrow peaks in between a zero at the double charm open channel.   

The strong phase variation is an important signature to be observed in both charm loops. In the partonic one the phase starts at zero in the double charm threshold, around 3 GeV, and rise abruptly after that.
In the hadronic one, the change of the phase sign, Fig.~\ref{fig:LoopH}(right), is placed in a region close  where data, Fig. \ref{lhcb} (right), shows a CP asymmetry change in sign.
Although we factorized the study of each charm loop, both are expected to contribute to the final amplitude. It is worth mention that we are not considering all the nonresonant nonperturbative sources. There could be other charm hadronic triangles with heavier mesons besides other source amplitudes such as the rescattering $\pi\pi \to KK$. Moreover, these nonresonant amplitudes are placed in a rich environment with other resonant amplitudes whose interference are not trivial. More than proving that the observed CP violation data is given by the specific hadronic loop described 
in Fig.\ref{fig:LoopH}, we provide one important final state interaction (FSI) mechanism which could produce CP asymmetry at higher energies.

To illustrate our discussion, we briefly recall previous CP violation studies\cite{BOT,ABAOT} where the leading order (LO) decay amplitude 
including the FSI, which respects the CPT constraint\cite{wolfenstein},  is written as:
\begin{equation}
\mathcal{A}^\pm_{LO} = A_{0\lambda} + e^{\pm i\gamma}B_{0\lambda}
+ i\sum_{\lambda^\prime}t_{\lambda^\prime,\lambda}\left(A_{0\lambda^\prime} + e^{\pm 
i\gamma}B_{0\lambda^\prime}\right),
\label{cp24}
\end{equation}
where $\gamma$ is the weak phase, the  amplitude source are represented by $A_{0\lambda}$ and $B_{0\lambda}$, 
$\lambda$ is the hadronic channels and 
$t_{\lambda^\prime,\lambda}=\imath\left(\delta_{\lambda^\prime,\lambda}-S_{\lambda^\prime,\lambda}\right)$ 
the scattering amplitude between channels $\lambda$ and 
$\lambda^ \prime$ coupled by the strong interaction S-matrix $\left(S_{\lambda^\prime,\lambda}\right)$. 

The leading order
decay amplitude Eq.(\ref{cp24}) can be put in correspondence with both the partonic Eq.(\ref{Ap}) and the hadronic loop Eq.(\ref{A1.2}).
In this case, the partonic loop is associated with $A_{0\lambda}$ and the hadronic loop with 
$\imath \,t_{\lambda^\prime,\lambda}\,A_{0\lambda^\prime}$, with the proviso that the $D\bar D$ 
in the hadronic loop is taken as on-mass-shell 
contribution. The source terms $B_{0\lambda}$ are the ones carrying the weak phase. 
The CP asymmetry  is given by $\Delta 
\Gamma_\lambda=|{A}^-_{LO}|^2-|{A}^+_{LO}|^2$, which leads to: 
\begin{small}
\begin{equation}
\Delta \Gamma_\lambda
=4(\sin\gamma) \, \mbox{Im}\Bigg\{ 
\left(B_{0\lambda}\right)^*A_{0\lambda} 
+i\sum_{\lambda^\prime}\left[\left(B_{0\lambda}\right)^*t_{\lambda^\prime, 
\lambda}\,
A_{0\lambda^\prime} - \left(B_{0\lambda^\prime}\, t_{\lambda^\prime,\lambda}\right)^*
A_{0\lambda}\right]\Bigg\},
\label{cp26}
\end{equation}
\end{small}
where in the right hand side, the second and third terms are associated with ``compound'' CP 
asymmetry~\cite{Soni2005}. Therefore, the interference between the source terms, the partonic loop and the ones 
carrying the FSI is evident and suggests that the position of the sign 
change in the CP asymmetry (see Fig. \ref{lhcb} right) can be shifted with respect to the sign change  position in the phase of the hadronic loop given
in Fig.\ref{fig:LoopH}.

In order to  evaluate our proposal, namely the relevant contribution of the hadronic loops and the partonic loop in different kinematic
regions, it is important that the future amplitude analysis of the $B^\pm \to K^\pm K^+ K^-$ decay include these amplitudes 
in their data fits. Only then we will be able to confirm the clear separation of the relevance of partonic vs hadronic loops considering the 
final state interaction.

 In summary, motivated by the separation of the short and long distance physics in the distribution of events in the Dalitz plane for
the $B^\pm \to K^\pm K^+ K^-$ decay, 
we invoke a hadronic description,  which we confirm that presents a very distinct pattern from the partonic 
one  in the allowed kinematic region, driven 
strongly by the final state interaction amplitude, which couples the virtual intermediate double charm state to 
the  $K^+K^-$  channel, and leaving a noticeable mark in the high mass region. Such mechanism could be important to explain the CP violation observed at high mass.

{\it Acknowledgements.} 
PCM would like to thanks Jean Marc Gerard for the fruitful discussion on the work. 
This work was partly supported by the Funda\c c\~ao de Amparo \`a Pesquisa do Estado de
 S\~ao Paulo [FAPESP grant no. 17/05660-0],
Conselho Nacional de Desenvolvimento Cient\'ifico e Tecnol\'ogico [CNPq grants no.
 308025/2015-6, 308486/2015-3 ] and
Coordena\c c\~ao de Aperfei\c coamento de Pessoal de N\'ivel Superior (CAPES) of Brazil.
This work is a part of the project INCT-FNA Proc. No. 464898/2014-5.
\appendix
\section{ S-matrix and scattering amplitude model}
\label{ap:DDKK}
The two channel S-matrix is parametrized as
\begin{equation}\label{sma}
S =  \left( \begin{array}{cc} \eta \, e^{2 i \delta_1} & i\sqrt{1-\eta^2} \, e^{i (\alpha + \beta)} \\
i\sqrt{1-\eta^2} \, e^{i (\delta_1 + \delta_2)} & \eta \, e^{2 i \delta_2} \end{array} \right)
\end{equation}
where $\delta_1$ and $\delta_2$ are the phase-shifts and $\eta$ is the inelasticity parameter, which accounts for the probability flux between the two coupled channels.
Since we are dealing with a three-body decay, the FSI effect will appear as a distribution depending on one of the two-body invariant masses, therefore the scattering amplitude cannot be obtained only asymptotically. We deal only with  the S-wave amplitude, while the amplitudes inspired in the Regge theory~\cite{Donoghue1996,Suzuki2008} needs to carry the dependence on higher angular momentum partial waves.

Our proposal for the off-diagonal matrix element is:
\begin{equation}\label{absorption}
\sqrt{1-\eta^2} = {\mathcal N} \sqrt{s/s_{th\,2}-1}\,\left({s_{th\,2}\over s}\right)^{\xi}
\end{equation}
where ${\mathcal N}$ is a normalization. For the phases we suggest the following parametrization:
\bea
e^{2 i \delta_1} &=& 1 - \frac{2 i k_1}{c + b\,k_1^2 +i k_1}={c + b\,k_1^2 -i k_1\over c + bk_1^2 +i k_1} 
\\[2mm]
e^{2 i \delta_2}&=& 1 - \frac{2 i k_2}{\frac{1}{a}+i k_2}={\frac{1}{a}-i k_2\over\frac{1}{a}+i k_2}
\eea
where $k_1=\sqrt{\frac{s-s_{th\,1}}{4}}$ and  $k_2=\sqrt{\frac{s-s_{th\,2}}{4}}$. For channel 2, we choose a scattering length dominated parametrization.
The scattering amplitude is defined as 
$t_{ij}=i(\delta_{ij}-S_{ij})$. 
Above the
 threshold, $s>s_{th2}$, the expression of $t_{12}$ become: 
\begin{eqnarray}\label{tij}
&&t_{12}= - i\,
\sqrt{1-\eta^2}\, \left[\left({c + bk_1^2 -i k_1\over c+bk_1^2+i k_1}\right)\,\,
\left({\frac{1}{a}-i k_2\over\frac{1}{a}+i k_2}\right)\right]^\frac12 \ .
\end{eqnarray}

\begin{figure}[ht]
\begin{center}
 \includegraphics[width=.35\columnwidth,angle=0]{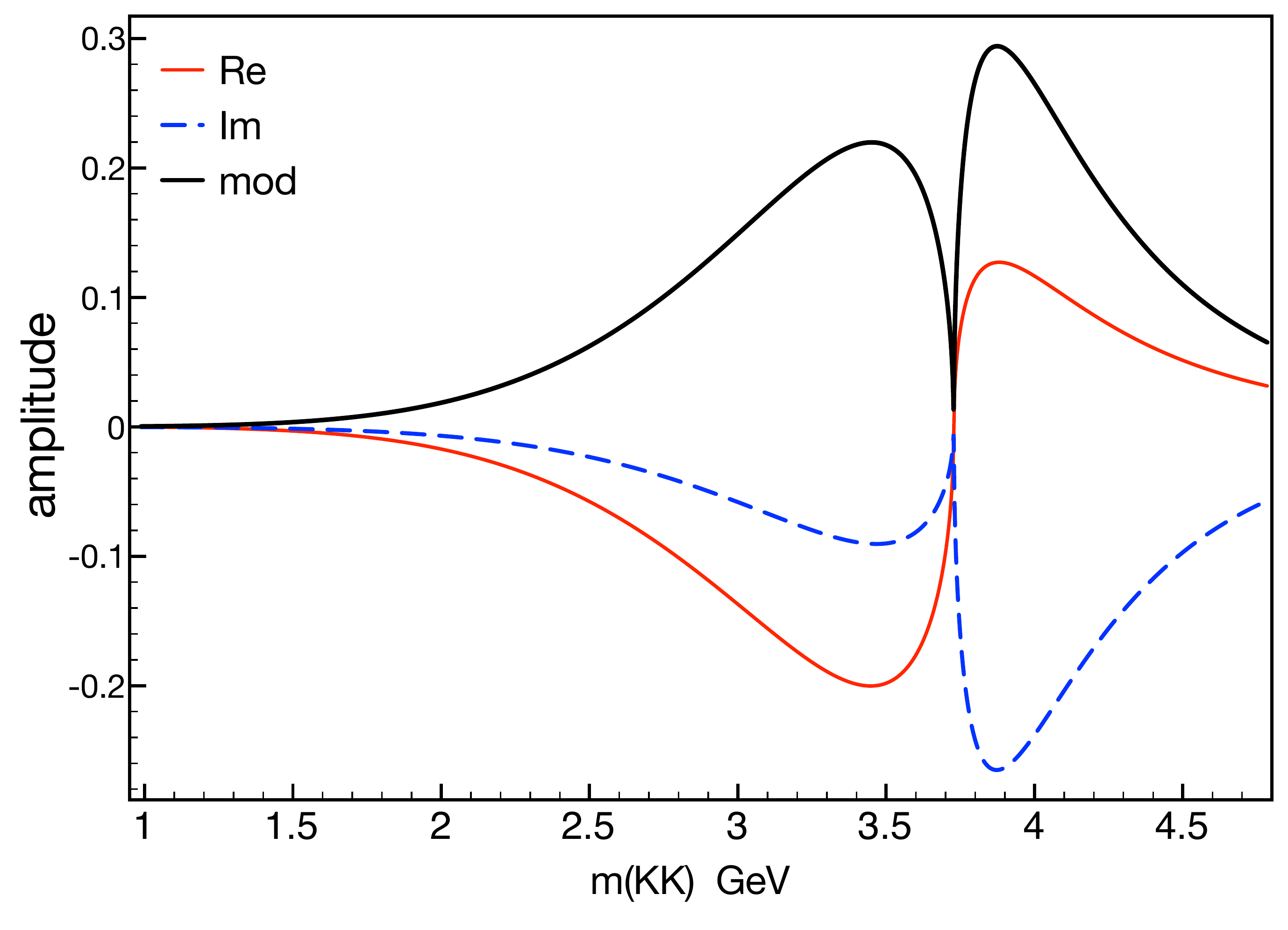}
 \hspace*{4mm}\includegraphics[width=.35\columnwidth,angle=0]{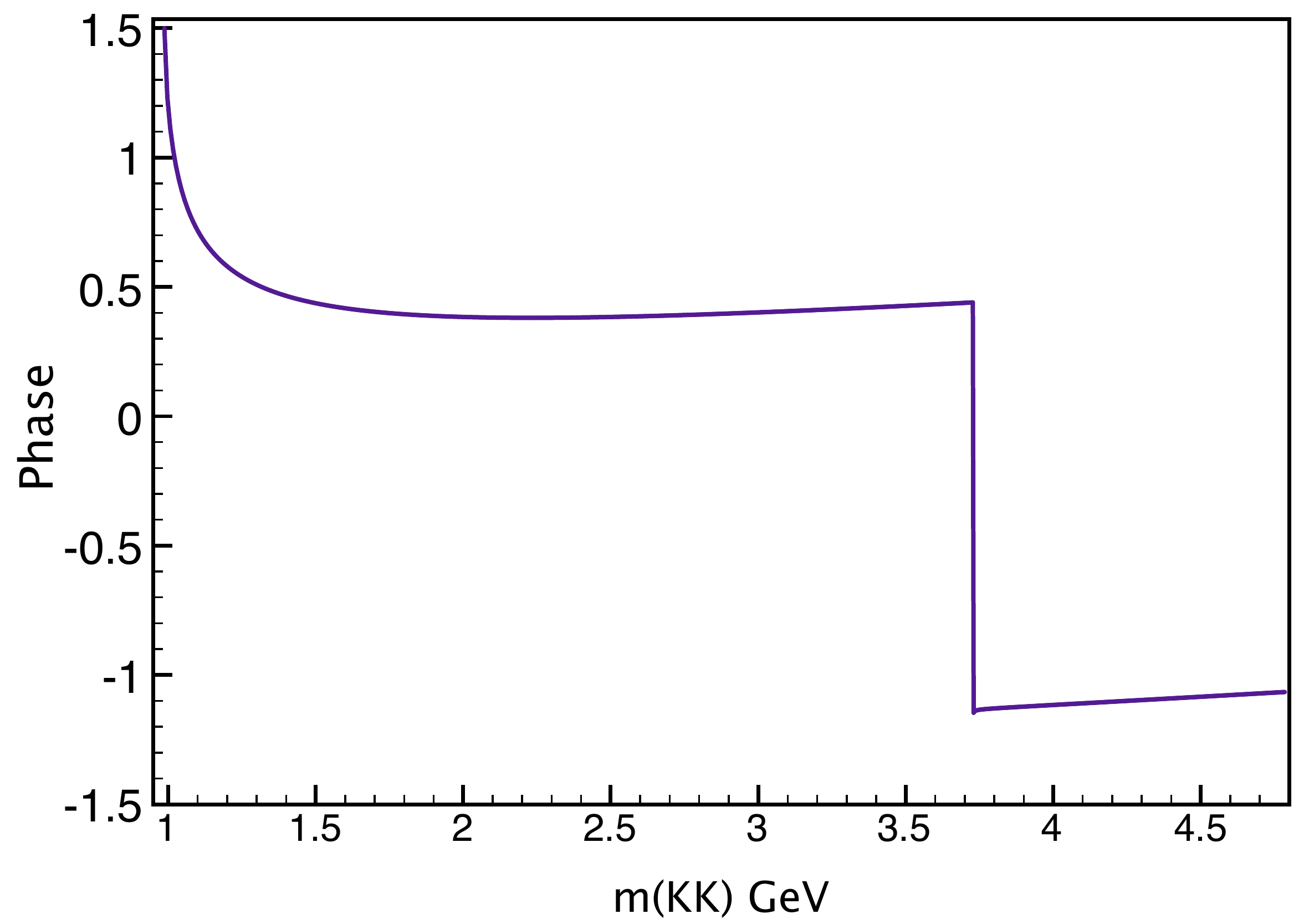}
\caption{ Amplitude for $D \bar{D}\to K\bar{K}$ scattering, eqs.(\ref{t12c}) and (\ref{t12d}): (left) modulus, real and imaginary parts;  (right) phase in radians. }
\end{center}
\label{fig3}
\end{figure}

The analytic continuation of the transition amplitude $t_{12}$ below the threshold of
channel 2, can be obtained noticing that $k_2\to i\kappa_2$ for $s<s_{th\,2}$, and now $\kappa_2=\sqrt{s_{th\,2}-s}/2$. 
However, one needs to take care of the amplitude behaviour at low values of $s$, once it modulus was tailored to reproduce power-law decrease at large momentum. One phenomenological 
possibility is to introduce an infrared cutoff in (\ref{absorption}) as follows:
\begin{eqnarray}\label{absorptionb}
\sqrt{1-\eta^2}=
{\mathcal N} (s/s_{th\,2})^\alpha\,\sqrt{s/s_{th\,2}-1}\,\left({s_{th\,2}\over s+s_{QCD}}\right)^{\xi+\alpha}
\end{eqnarray}
where $s_{QCD}$ is an infrared cut-off estimated to be of the order of the hadronic scale $s_{QCD}\sim 1$ GeV$^2$. In addition, we introduce a factor $s$ in the non-physical region, expressing that 
the coupling between the open channel of the two light-quarks and the closed channel 
of the two-heavy quarks is damped when entering deeply in the non-physical region as $s^\alpha$. Note
that we have kept the asymptotic power of the amplitude, namely $\sim s^{-\xi}$. 
Therefore, our proposal for the scattering amplitude $D\bar{D}\to K\bar{K}$ and the analytic continuation below 
threshold, $s< s_{th\,2}$, is given by:
\begin{equation} 
t_{12}= \,
{\mathcal N}\, \frac{s^\alpha}{s_{th\,2}^\alpha}\, \frac{2\kappa_2}{\sqrt{s_{th\,2}}}\,\left({s_{th\,2}\over s+s_{QCD}}\right)^{\xi+\alpha}
\left[\left({ c + bk_1^2-i k_1 \over c + bk_1^2 +i k_1}\right)\,\,
\left({\frac{1}{a}+ \kappa_2\over\frac{1}{a}-\kappa_2}\right)\right]^\frac12\, ,
\label{t12c}
\end{equation}
and for $s\geq s_{th\,2}$ is written as:
\begin{equation} 
t_{12}= \, - i\,
{\mathcal N} \,\frac{2\,k_2}{\sqrt{s_{th\,2}}}\,\,\left({s_{th\,2}\over s+s_{QCD}}\right)^{\xi}\, \left({m_0\over s- m_0}\right)^{\beta} 
\left[\left({\frac{c}{1-s/s_0}-i k_1\over \frac{c}{1-s/s_0}+i k_1}\right)\,\,
\left({\frac{1}{a}-i k_2\over\frac{1}{a}+i k_2}\right)\right]^\frac12\, , \label{t12d} 
\end{equation}
where $\left({m_0\over s- m_0}\right)^{\beta}$ was introduced to modulate the shape of the amplitude bump.  

The parameters should be fitted to the data. But, in order to produce a toy Monte Carlo for the transition amplitudes (\ref{t12c}) and (\ref{t12d}) we guessed them following the phenomenology inputs. The parameter  $b$ and $c$ are residues of the pole in 
$k\cot \delta$ expression and we used $c=0.2$ and $b=1$. For the scattering length $a$ in the 2 channel, we can take the limiting case $a\to \pm \infty$, namely the two heavy mesons are 
strongly interacting close to the threshold.  The IR scale $s_{QCD}$ is of the order of 1 GeV$^2$, or may be less $\sim \Lambda_{QCD}^2$ and  from previous studies~\cite{CPV} we found $\xi\sim 2.5$. For the other ad doc parameter we chose: $\alpha =3$, but higher powers are not excluded, $m_0=8$ and $\beta=2$.  With this choice of parameter our scattering amplitude is given in Fig.~\ref{fig3}.

\end{document}